\documentstyle[12pt]{article}
\def\beq{\begin{equation}}
\def\brr{\begin{array}}
\def\err{\end{array}}
\def\eeq{\end{equation}}
\def\bea{\begin{eqnarray}}
\def\eea{\end{eqnarray}}
\def\bs{\bigskip}
\def\ni{\noindent}

\def\nn{\nonumber}
\def\ms{\medskip}

\begin{document}

\hfill UB-ECM-PF 92/7
\mbox{}

\vspace*{1cm}

\begin{center}

{\LARGE \bf
Renormalization group equations in curved space-time with non-trivial
topology}

\vspace{1cm}

{\sc E. Elizalde and S.D. Odintsov}\footnote{On leave from Department of
Mathematics and Physics,
 Pedagogical Institute, 634041 Tomsk, Russia.}
\ms

Department E.C.M., Faculty of Physics, \\
University of Barcelona, \\
Diagonal 647, 08028 Barcelona, Spain \\
{\it e-mail: eli @ ebubecm1.bitnet}

\vspace{1cm}

{\sl March 1992}

\vspace{2cm}

{\bf Abstract}

\end{center}

Renormalization group equations for massless GUT's in curved
space-time with non-trivial topology are formulated. The asymptotics of
the effective action both at high and low energies
are obtained. It is shown that the Casimir energy contribution at high
curvature
(early Universe) becomes non-essential in the effective action.

\vspace{1.5cm}


\newpage

The investigation of the different aspects of quantum field theory in
spaces with non-trivial topology (see [1-3] and references therein) is
interesting for several reasons. First, it is possible that the
early Universe was in fact multiconnected. Second, in the Kaluza-Klein
approach some of the dimensions of the Universe must be compactified
(what provides the space with a non-trivial topology). Third,  field
theory in
the simplest space with non-trivial topology (the one-torus compactified
space) is quite similar to field theory at non-zero temperature.

Beautiful realizations of the Casimir effect corresponding to
different field
theories on spaces with non-trivial topology have been investigated in
a number of works (see [4], for instance, and references therein). The
question of
the definition of the vacuum energy is a most fundamental one in any
quantum
field theory, and it transforms into a highly involved problem when
curved space, general boundaries, and non-trivial topologies are
considered. Very skillful regularizations have to be used and these
studies become a playground for various elegant general methods of
analysis.

So, the different ingredients (and outcoming results) of quantum field
theory
in curved space-time with non-trivial topology are highly interesting.

In the present letter, we formulate the renormalization group equations
in curved space with non-trivial topology. For simplicity, we restrict
ourselves to the particular space $R_3 \times S_1$, where $R_3$ is a
curved three-dimensional space while $S_1$ is the one-dimensional torus.
We shall calculate the asymptotics of the effective action for an
arbitrary asymptotically-free GUT defined on such a background.
In particular, we are going to construct the explicit general form of
the renormalization group equations, what will allow us to
investigate the different asymptotics (high and low energies) of the
effective action. From such investigation carried out for asymptotically
free GUTs we will conclude that ---for these theories--- the
consideration of the topological
Casimir energy is actually not essential in the early Universe (strong
curvature limit).
\bs

Let us start from a massless GUT which includes scalars, spinors
and gauge fields. Denote the set of fields by $\varphi$ and the
set of coupling constants by $f$. The multiplicatively
renormalized Lagrangian has the following schematical form (see
[5,6] for a review and details)
\bea
{\cal L} &=& \sqrt{-g} \ ({\cal L}_{ext} + {\cal L}_m ) , \nn \\
{\cal L}_{ext} &=& aR^2 +b  \, G +c \, C^2_{\mu \nu \alpha \beta} + d \,
\Box R, \\
{\cal L}_m &=& -\frac{1}{4} G_{\mu \nu}^a G^{a\mu\nu} + \frac{1}{2} (
D_{\mu} \Phi )^2 + \frac{1}{2} \xi R \Phi^2
+ \bar{\psi} \not\mbox{D} \, \psi + h \bar{\psi} \psi \Phi - \lambda
\Phi^4. \nn
\eea
Here $\varphi = \left\{ A^a_{\mu}, \psi, \Phi \right\}$ is the set of
fields and $ f = \left\{ g,h,\lambda, \xi, a,b,c,d \right\}$ is the set
of dimensionless coupling constants.

The renormalization group equations in curved space-time (for a
review see [5,6]) follow from the condition of multiplicative
renormalizability. In the case under consideration, we
get\footnote{In the non-zero temperature formulation, $L$ can be
considered as the inverse temperature.}
\beq
\Gamma  (g_{\mu\nu} , L, \varphi_0, f_0 ) =  \Gamma  (g_{\mu\nu}
, L, \varphi, f, \mu ),
\eeq
where $L=2\pi R$, $R$ being the radius of $S_1$ and $\mu$ the
mass parameter. Note that what we have in mind at this point is
dimensional
regularization and  the scheme of minimal substraction. Then the
renormalization constants and $\beta$-functions do not depend on
$L$.

It follows from (2), that
\beq
\left( \mu \frac{\partial}{\partial \mu} + \beta_f
\frac{\partial}{\partial f} + \gamma \varphi \frac{\delta}{\delta
\varphi} \right)   \Gamma  (g_{\mu\nu} , L, \varphi, f, \mu ) =0,
\eeq
where $\beta_f$ is the $\beta$-function and  $\gamma$ the
$\gamma$-function, and $ \varphi \frac{\delta}{\delta \varphi}
\equiv \int d^4x \,  \varphi \frac{\delta}{\delta \varphi}$.

Making use of the scale transformation of the metric: $g_{\mu\nu}
\rightarrow e^{-2t} g_{\mu\nu}$, we know that
\beq
 \Gamma  (e^{-2t} g_{\mu\nu} , e^{-t} L, e^{d_{\varphi} t} \varphi, f,
e^t \mu ) =  \Gamma  (g_{\mu\nu} , L, \varphi, f, \mu ).
\eeq
With this transformation, eq. (4) results into the following
\bea
\left( D - L \frac{\partial}{\partial L}  \right)
 \Gamma  (e^{-2t} g_{\mu\nu} ,  L,  \varphi, f,  \mu ) & = & 0,
\mbox{\hspace*{6cm}}  (5a) \nn \\
D \, \Gamma  (e^{-2t} g_{\mu\nu} , e^{-t}  L,  \varphi, f,  \mu ) &
= & 0,  \mbox{\hspace*{6cm}}  (5b) \nn \\
\left( D - 2 g_{\mu\nu}\frac{\delta}{\delta g_{\mu\nu}}  \right)
 \Gamma  (g_{\mu\nu} , e^{-t}  L,  \varphi, f,  \mu ) & = & 0,
\mbox{\hspace*{6cm}} (5c) \nn
\eea
where
\[
D=  \mu \frac{\partial}{\partial \mu} + \frac{\partial}{\partial
t} + d_{\varphi}  \varphi \frac{\delta}{\delta \varphi}.
\]
To our knowledge, these renormalization group equations (5), which
correspond
to curved space-time with non-trivial topology,  have never been
derived
before in the literature. Combining (5a)-(5c) with (3), we can get rid
of $ \mu
\frac{\partial}{\partial \mu}$ and thus obtain three different forms for
the renormalization group equation in curved space with non-trivial
topology. These are the following:
\bea
\left( D_t - L \frac{\partial}{\partial L}  \right)
 \Gamma  (e^{-2t} g_{\mu\nu} ,  L,  \varphi, f,  \mu ) & = & 0,
\mbox{\hspace*{6cm}} (6a) \nn \\
D_t \, \Gamma  (e^{-2t} g_{\mu\nu} , e^{-t}  L,  \varphi, f,  \mu )
& = & 0,  \mbox{\hspace*{6cm}} (6b) \nn \\
\left( D_t - 2 g_{\mu\nu}\frac{\delta}{\delta g_{\mu\nu}}
\right)
 \Gamma  (g_{\mu\nu} , e^{-t}  L,  \varphi, f,  \mu ) & = & 0,
\mbox{\hspace*{6cm}} (6c) \nn
\eea
where
\[
D_t= \frac{\partial}{\partial t} - \beta_f
\frac{\partial}{\partial f}  - \left(\gamma - d_{\varphi}\right)
\varphi \frac{\delta}{\delta \varphi}.
\]
Equations (6) constitute the main result of the present letter.

The solutions of eqs. (6a)-(6c) are
\bea
 \Gamma  (e^{-2t} g_{\mu\nu} ,  L,  \varphi, f,  \mu ) & = &
\Gamma  ( g_{\mu\nu} ,  L_0 e^t,  \varphi (t), f(t),  \mu ) ,
\mbox{\hspace*{5cm}} (7a) \nn \\
 \Gamma  (e^{-2t} g_{\mu\nu} , e^{-t}  L,  \varphi, f,  \mu ) & =
& \Gamma  ( g_{\mu\nu} ,
  L,  \varphi (t), f(t),  \mu ) ,  \mbox{\hspace*{5.5cm}}
(7b) \nn \\
 \Gamma  (g_{\mu\nu} , e^{-t}  L,  \varphi, f,  \mu ) & = &
\Gamma  ( g_{\mu\nu}^0 e^{2t} ,  L,  \varphi (t), f(t),  \mu ) ,
\mbox{\hspace*{5cm}} (7c) \nn
\eea
where
\[
\frac{df(t)}{dt}= \beta_f (t), \ \ f(0)=f, \ \ \ \ \ \
\frac{d\varphi (t)}{dt}= (\gamma (t)- d_{\varphi} ) \varphi (t),
\ \  \varphi (0) = \varphi ,
\]
and $L_0$ and $g^0_{\mu \nu}$ are the values of $L$ and $g_{\mu
\nu}$ at $t=0$, respectively.
When $g_{\mu\nu} \rightarrow e^{-2t} g_{\mu\nu}$, then $R^2
\rightarrow e^{4t} R^2$ (for a general discussion see [5,6]).
Here, equation (7a) gives the possibility to investigate the
theory in the strong curvature limit (high energies) when $t
\rightarrow \infty$. On the other hand, for equation (7b),  $t
\rightarrow \infty$ corresponds to the strong curvature and small
radii  limit. Finally, for equation (7c) the limit $t \rightarrow
\infty$ corresponds to the small radii limit (infrared limit).
This follows from the fact that, on the left hand side of (7c),
curvature does not change, while on the right hand side of this
equation, when $t\rightarrow \infty$ then $R^2 \rightarrow e^{-4t} R^2$,
i.e., the limit $t\rightarrow \infty$ is the small curvature (or low
energy, or big distance) limit, namely, the infrared limit.

One can now look to the asymptotics of the effective action for
GUT's like (1). In the case of {\sl asymptotically free GUT's}
only the limit $t \rightarrow +\infty$ exists (not $t \rightarrow -
\infty$). The asymptotic of the effective action is given by the
classical action, with coupling constants and fields changed by
the effective coupling constants and effective fields,
respectively. (This picture follows from asymptotic freedom).
{}From now on, for simplicity, we shall put $A_{\mu}^a$ and $\psi$
equal to zero. We shall only consider scalar and gravitational
backgrounds.

The typical behaviour of the effective coupling constants and
fields for GUT's is [5,6,7] ($t \rightarrow \infty$)
\setcounter{equation}{7}
 \[
g^2(t) \sim \frac{b}{t}, \ b>0, \ \ \ \ \lambda (t) \sim g^2(t),
\ \ \ \ h^2(t) \sim g^2(t),\]
\beq
\xi (t) \sim \frac{1}{6} + \left( \xi - \frac{1}{6} \right)
t^{a_{\xi}}, \ \ \ \ \Phi (t) \sim \Phi e^{-t} t^{-
a_{\Phi}b},
 \eeq
where for different models, the constants $a_{\xi}$ and
$a_{\Phi}$ can be  positive or negative. The effective
coupling constants, $a(t), b(t), c(t)$ and $d(t)$ behave, as a
rule, as $\sim const \times t$, when  $t \rightarrow \infty$.

Substituting  (8) in (7a)  we get that the asymptotics of the
effective action are given by ${\cal L}_{ext}$, as in the case of a
space with trivial topology [5,6]. (Normally, the term in the square of
the Weyl tensor is the leading one among the terms of ${\cal L}_{ext}$).
The same is true for equation
(7b). However, for equation (7c) the situation is quite different.

Here the analysis must take into account the finite topological
correction to the effective action provided by the Casimir
energy, i.e.
\beq
\Gamma_C = \int d^4x \, \sqrt{g} \, \frac{C}{L^4},
 \eeq
 where the constant $C$ has the form [8]:
\beq
C= \lambda \, \left( N_{\varphi} + c_{1/2} N_{\psi} + c_1 N_A
\right).
\eeq
The $N$'s mean here number of fields of each given type, while the
constant $\lambda$ and  the coefficients $c$'s only depend on the number
of compactified dimensions (assuming periodic boundary conditions) or on
the kind of specific boundary conditions imposed. In particular, for a
single compactified dimension, we have
\beq
\lambda = - \frac{\pi^2}{90}
\eeq
and the coefficients are simply the spin multiplicity factors (negative
for fermions, so $c_{1/2} =-4$). One must observe that these are the
only `boundary conditions' which are compatible for all kind of spins,
and correspond to Kaluza-Klein theories. If, on the other hand, we
impose the `ordinary' conditions of the (electromagnetic) Casimir
effect ---i.e. those corresponding to perfectly conducting parallel
plates, related with Dirichlet boundary conditions for the scalar
fields--- then the value of $\lambda $ is given by
\beq
\lambda = - \frac{\pi^2}{1440}
\eeq
and moreover $c_1 = 2$ [8].\footnote{Notice again that Dirichlet
boundary
conditions for the scalar field do not admit a generalization to fields
with spin.} Finally, the same coefficients are valid in both cases when,
instead of one, two of the dimensions are compactified.

Summing up, we see that the leading part of the effective action
when $t \rightarrow \infty$ is
\beq
\Gamma \sim \int d^4x \, \sqrt{g} \, e^{3t} \, \frac{C}{L^4},
\eeq
As a consequence, the topological correction to the classical
action (1) actually becomes important in the small radii limit (at some
fixed curvature), i.e. in the infrared limit.

Our investigation shows that if one describes the early Universe
in the GUT's epoch by means of some model of asymptotically free
GUT, like $SU(5)$ or $E_6$, one can consistently neglect the
topological corrections to the effective action at strong
curvature (near the Planck scale). In such class of theories the
Casimir effect is not important, as we have argued, for
considerations about the early Universe. However, the situation
changes completely if we describe the early Universe with the help of a
 Kaluza-Klein type of theory.
Indeed, because of the non-renormalizability of Kaluza-Klein theories,
the above renormalization group equations are not valid. From the other
side, if we identify $L$ with the radius of the extra dimension
(consider $d=5$, for simplicity), to be consistent with the
observations it is well known that we need to take $L\sim
L_{\mbox{Planck}}$. Here the Casimir energy gives the biggest
contribution to $\Gamma \sim L^{-4}_{\mbox{Planck}}$, and is expected to
be very important.

To conclude, it may be interesting to note that this asymptotic
effective action does not change in $R^2$-gravity interacting
with GUT matter (see [5,6] for a review). Again, the Casimir
energy is essential in equation (7b), where we should include
into the constant $C$ of (13) the contribution from the quantum
gravitational field (of course, in this case the asymptotic
behaviours (8) are different). Finally, the extension of these
considerations to other spaces ---in particular with non-trivial
topology (as hyperbolic spaces)--- is straightforward.

\vspace{2cm}

\ni{\large \bf Acknowledgments}

Discussions with T. Muta and H. Osborn  are greatly
appreciated. S.D.O. thanks the members of the Department E.C.M. of
Barcelona University for the kind hospitality.
E.E. thanks the Alexander von Humboldt Foundation for continued help.
The comments of the referees are also appreciated.
This work has
been supported by Direcci"n General de Investigaci"n
Cient!fica y Tcnica (DGICYT), research projects
 PB90-0022 and SAB92-0072.


\newpage

\begin{thebibliography}{99}


\bibitem{} B.S. De Witt, C.F. Hart and C.J. Isham, {\sl Physica}
{\bf A96} (1979) 197; J.S. Dowker and R. Banach, {\sl J. Phys.}
{\bf A1} (1978) 2255.

\bibitem{}  C.J. Isham, {\sl Proc. R. Soc.} {\bf A362} (1978)
383; A. Chockalingham and C.J. Isham, {\sl J. Phys.} {\bf A13}
(1980) 2723.

\bibitem{}  G. Kennedy, R. Critchley and J.S. Dowker, {\sl Ann. Phys.
(N.Y.)}, {\bf 125} (1980) 346.
 Yu.P. Goncharov and A.A. Bytsenko, {\sl
Class.
Quant. Grav.} {\bf 4} (1987) 555; I.L. Buchbinder and S.D. Odintsov,
{\sl Fortschr. Phys.} {\bf 37} (1989) 225.

\bibitem{}
 J.S. Dowker and G. Kennedy, {\sl J. Phys.} {\bf A11} (1978) 895;
 S.K. Blau, M. Visser and A. Wipf, {\sl Nucl. Phys.} {\bf B310} (1988)
163;
E. Elizalde and A. Romeo, {\sl Rev. Math. Phys.} {\bf 1} (1989) 113;
E. Elizalde and A. Romeo, {\sl Int. J. Mod. Phys.} {\bf A5} (1990) 1653;
K. Kirsten, {\sl J. Math. Phys.} {\bf 32} (1991) 3008;
A.A. Bytsenko and S. Zerbini, {\sl The Casimir effect for a class of
hyperbolic D+1 dimensional space-times}, preprint Universit di Trento.

\bibitem{} I.L. Buchbinder, S.D. Odintsov and I.L. Shapiro, {\sl
Effective Action in Quantum Gravity}, A. Hilger, Bristol, U.K., 1992.

\bibitem{} I.L. Buchbinder, S.D. Odintsov and I.L. Shapiro, {\sl
Rivista Nuovo Cim.} {\bf 12} (1989) 1.

\bibitem{} T. Muta and S.D. Odintsov, {\sl Mod. Phys. Lett.} {\bf
A6} (1991) 3641.

\bibitem{} N.D. Birrell and P.C.W. Davies, {\sl Quantum fields in curved
space}, Cambridge University Press, Cambridge, U.K., 1982;
J. Ambj{\o}rn and S. Wolfram, {\sl Ann. Phys. (N.Y.)} {\bf 147} (1983)
1;
E. Elizalde, {\sl Nuovo Cim.} {\bf 104B} (1989) 685;
E. Elizalde and A. Romeo, {\sl Phys. Rev.} {\bf D40} (1989) 436.

\end{thebibliography}
\end{document}